\documentclass{article}

\usepackage{arxiv}

\usepackage[utf8]{inputenc} % allow utf-8 input
\usepackage[T1]{fontenc}    % use 8-bit T1 fonts
\usepackage{hyperref}       % hyperlinks
\usepackage{url}            % simple URL typesetting
\usepackage{booktabs}       % professional-quality tables
\usepackage{amsfonts,amsmath}
\usepackage{nicefrac}
\usepackage{microtype}
\usepackage{graphicx}
\usepackage{bm}
\usepackage{amssymb}
\usepackage{array}
\usepackage{mathtools}
\usepackage{xcolor}
\usepackage{tabularx}
\usepackage{algorithm,algorithmicx,algpseudocode}
\usepackage{enumitem}
\usepackage{float}
\usepackage[square,comma,numbers]{natbib}
\usepackage{multicol}

\definecolor{mred}{RGB}{0,0,0}
\definecolor{mblue}{RGB}{57,106,177}
\hypersetup{
    citecolor=blue,
    colorlinks=true,
    linkcolor=blue,
    filecolor=blue,
    urlcolor=blue
}

\graphicspath{{Figures/},{./},{images/}}
\title{Modelling non-linear aeroelastic loads in long-span bridges with extreme learning machines}

\author{
  Gledson Rodrigo Tondo$^{\star}$ \\
  Bauhaus-Universit\"at Weimar\\
  Weimar, Germany \\
  \And
  Samir Chawdhury \\
  Bauhaus-Universit\"at Weimar\\
  Weimar, Germany \\
  \AND
  Sergio Andres Castro Giraldo \\
  Bauhaus-Universit\"at Weimar\\
  Weimar, Germany \\
  \And
  Guido Morgenthal \\
  Bauhaus-Universit\"at Weimar\\
  Weimar, Germany \\
}

\begin{document}
\maketitle
\begin{abstract}
Accurate modelling of aerodynamic loads is essential for predicting instabilities and ensuring the safety of long-span bridges. A methodology is introduced for modelling aerodynamic self-excited forces in bridge-deck cross-sections using extreme learning machines (ELMs). ELMs, as single-layer feedforward neural networks, offer efficient training and accurate predictions. Forced-oscillation datasets from computational fluid dynamics (CFD) or wind-tunnel experiments are used for training, enabling systematic data selection to capture non-linear aerodynamic behaviour often missed by semi-analytical approaches. Once trained, the model predicts self-excited loads for any arbitrary motion composed by frequencies and amplitudes within the training domain. Comparisons with analytical, semi-analytical, and CFD results show superior accuracy in capturing non-linear force components and close agreement for aerodynamic loads and flutter wind speeds. Training required about 1.1\% of the time of a conventional neural network, and coupled flutter analysis runs in seconds, providing orders-of-magnitude speed-ups over CFD. These results indicate that ELM-based frameworks are accurate, practical, and efficient alternatives for modelling self-excited loads, particularly when preliminary CFD or wind-tunnel data are available. The presented approach offers a reliable data-driven technique for aeroelastic load modelling in long-span bridges.
\end{abstract}


\section*{Notation}

\begin{multicols}{2}
	\begin{tabbing}
		\hspace{1.8cm} \= \kill
		$\mathcal{D} = \lbrace \bm{X}, \bm{y} \rbrace$
		\> Training dataset (inputs and targets) \\
		$\bm{o}$            \> ELM outputs \\
		$\bm{H}$            \> Hidden neural values \\
		$\bm{W}$, $\bm{b}$  \> Randomised weight matrix and bias vector \\
		$f(\cdot)$          \> Activation function \\
		$\bm{U}$, $\bm{\Sigma}$, $\bm{V}$
		\> Singular value decomposition matrices \\
		$\hat{\bm{\beta}}$  \> Optimal ELM weight matrix \\
		$\lambda$           \> Tikhonov regularisation parameter \\
		$L$                 \> Lift force \\
		$M$                 \> Moment \\
		$C_L$               \> Coefficient of lift \\
		$C_M$               \> Coefficient of moment \\
		$h$,$\dot{h}$,$\Ddot{h}$
		\> Vertical displacement, velocity and acceleration \\
		$\alpha$,$\dot{\alpha}$,$\Ddot{\alpha}$
		\> Angular displacement, velocity and acceleration \\
		$\rho$               \> Air density \\
		$U$                  \> Wind speed \\
		$B$                  \> Cross-section chord width \\
		$v_r$                \> Reduced velocity \\
		$f$                  \> Oscillation frequency \\
		$K$                  \> Non-dimensional frequency \\
		$H^*_{i}$, $A^*_{i}$
		\> Scanlan's derivatives ($i = \lbrace 1,\dots,4 \rbrace$) \\
		$Re$                 \> Reynolds number \\
		$m_h$, $m_{\alpha}$  \> Vertical and rotational masses \\
		$f_h$, $f_{\alpha}$  \> Vertical and rotational natural frequencies \\
		$\zeta$               \> Damping ratio \\
		$t$                   \> Time \\
		$U_{cr}$, $f_{cr}$   \> Critical flutter wind speed and frequency \\
		$N_t$                 \> Number of training points \\
		$N_d$                 \> Number of dimensions \\
		$N_n$                 \> Number of network neurons \\
		$N_c$                 \> Number of oscillation cycles \\
		$N_{tp}$              \> Number of training points per cycle \\
	\end{tabbing}
\end{multicols}

\section{Introduction} \label{sec:intro}

The study of self-excited forces in bridge aerodynamics is pivotal for ensuring the safety and performance of long-span bridges, especially under dynamic environmental conditions. These forces, which arise from complex interactions between the structural response and the - possibly laminar - airflow, can lead to flutter, an instability phenomenon that poses significant structural risks. As climate change induces extreme weather patterns, the potential for higher wind speeds further underscores the importance of accurately modelling self-excited forces~\citep{Intr_1,Intr_2,chu2021probabilistic}. Understanding and predicting these forces is essential for designing bridges that can withstand evolving environmental challenges while maintaining safety and durability.

Traditional modelling of self-excited forces on bridges relies on the linear framework introduced by~\cite{scanlan1971airfoil}, which represents aerodynamic drag, lift, and moment in the frequency domain using non-dimensional parameters called aerodynamic derivatives. These parameters are typically identified through computational fluid dynamics (CFD) or wind tunnel tests~\citep{sarkar2009comparative,diana2004forced,abbas2017methods}, and their values are highly sensitive to the dataset used for identification~\citep{abbas2016framework}. This sensitivity, combined with the linear nature of the model, introduces significant variability in estimating self-excited forces, especially under critical instability conditions such as flutter~\citep{abbas2016framework,cheng2005probabilistic,jakobsen2003modelling}. To address these limitations, various nonlinear aerodynamic load models have been proposed, including quasi-steady models and their extensions~\citep{kovacs1992analytical,diana2010aerodynamic,diana2004forced} and hybrid or multimode nonlinear models~\citep{chen2001nonlinear, chen200017,MATSUMOTO1997871}. Furthermore, rheological models~\citep{diana2008new}, hysteretic models~\citep{wu2013bridge}, and Volterra series approaches~\citep{skyvulstad2021use,skyvulstad2023regularised} have been used to model nonlinearities observed in self-excited forces for cases of bluff bodies or twin deck girders~\citep{GAO2018227,ZHOU20191072}. Approaches to reconstruct aerodynamic loads from structural responses have also been studied in recent years~\citep{tondo2023physics,tondo2024bayesian}. Recent advancements have introduced data-driven techniques to capture nonlinearities, employing methods such as deep neural networks~\citep{abbas2020prediction,CHEN20081925,CHEN2023105374}, support vector machines~\citep{LUTE2009830}, surrogate models~\citep{VERMA2024105769}, cellular automata networks~\citep{wu2011modeling}, long short-term memory (LSTM) neural networks~\citep{li2020novel,ZHAO2024105905}, and more recently transformers~\citep{montoya2024}. While effective, these methods can be computationally intensive and are sensitive to input parameters and modelling assumptions, as illustrated in~\citep{Hu2020NNbased,CASTELLON2021104484}. Novel methodologies based on physics-informed machine learning are promising tools to overcome these issues, but require a physical description of the phenomenon to be modelled, and are not straightforward to implement~\citep{kavrakov2022data,KAVRAKOV2024105848,TONDO2023103534,tondo2025efficient}.

To address these key challenges in modelling aerodynamic self-excited forces, this study introduces a data-driven framework based on extreme learning machines (ELMs)~\citep{huang2006extreme}. ELMs are single-layer feedforward neural networks known for their rapid training process, achieved by randomly assigning input weights and biases and using numerical matrix pseudo-inversion techniques. This eliminates iterative optimization, resulting in high computational efficiency while maintaining accuracy in capturing complex relationships, as demonstrated in numerous applications~\citep{liu2014extreme,ding2014extreme,wang2022review,ertugrul2016forecasting, ELM_APPLT_BEAM, ELM_APPLT_DAM}. The framework accelerates training and significantly reduces the computational cost of force predictions compared to traditional methods like CFD. In addition, it allows the incorporation of CFD or wind tunnel test results as training data, providing a physically consistent description of the aerodynamic loads. The model's data-driven nature allows it to retain the nonlinear properties of the training data, overcoming the simplifications inherent in conventional semi-analytical models. Additionally, the use of simplified forced vibration datasets for training reduces the reliance on extensive CFD or experimental campaigns, as such data are typically available in early design stages from preliminary wind tunnel tests or simplified CFD simulations. Since the forced vibration data capture a wide range of oscillation frequencies and amplitudes, the trained ELM can be extended to predict coupled structural–aerodynamic responses. This renders the proposed framework a versatile and efficient tool for investigating complex aeroelastic phenomena in bridge aerodynamics, while circumventing the need for costly or time-consuming simulations.

This paper is organized as follows: Section~\ref{sec:2} outlines the primary objectives of the proposed framework, introduces the theoretical foundations of ELMs, and explains their application to modelling aeroelastic forces. It also describes the training and prediction methodology using ELMs. Section~\ref{sec:flatplate} applies the framework to the analytical case of a flat plate model, serving as a benchmark to demonstrate ELM prediction capabilities and potential applications in bridge aerodynamics. Section~\ref{sec:bridgeaerodyn} extends the framework to the Great Belt East Bridge, focusing on nonlinear load components at high angles of attack, the accuracy of high-amplitude load predictions, and coupled flutter behaviour. Finally, Section~\ref{sec:conclusion} summarizes the key findings and conclusions. The implementation of the ELM model along with the applications in bridge aerodynamics can be found at \url{https://github.com/gledsonrt/SelfExcitedELMs}.

\section{ELM models for aerodynamic self-excited forces} \label{sec:2}

\subsection{Problem statement}

In bridge aerodynamics, various models are used to translate structural dynamic movement into equivalent dynamic loads~\citep{kavrakov2018synergistic}. These models are typically categorized into three main groups: semi-analytical, numerical, and experimental. Semi-analytical methods are generally computationally efficient but require inputs that often need to be obtained through CFD or wind tunnel testing. Their accuracy depends on the assumptions made during their derivation~\citep{kavrakov2019categorical}. Numerical methods, such as the finite element method, finite volume method, and vortex particle method, directly solve the fluid domain and its interaction with the structure using CFD. While more precise than semi-analytical approaches, numerical methods are computationally intensive and demand significant simulation time. Experimental methods involve constructing and testing scaled physical models in wind tunnels. This approach is considered the most reliable but is time-consuming and resource-intensive.

To bridge the gap between these model groups, the framework proposed in this work employs a data-driven approach to:

\begin{enumerate}
    \item Enhance the accuracy of semi-analytical models using training data from CFD simulations or wind tunnel analysis, by capturing nonlinear and fluid memory effects;
    \item Lower computational complexity (for CFD) and reduce time and resource costs (for experimental methods) by replicating their outputs with an efficient and simple data-driven technique;
    \item Leverage simplified forced vibration analysis data for training, enabling the use of basic CFD/experimental simulations and extending the model's capabilities to coupled structural-aerodynamic simulations.
\end{enumerate}

To this end, Section~\ref{sec:elmtheory} provides an overview of extreme learning machines and their extension to include regularization schemes, required for noisy datasets and to address limitations of the basic ELM model, while Section~\ref{sec:elmapplication} defines the approach to apply them to self-excited forces in bridge aerodynamics.

\subsection{Extreme learning machine models} \label{sec:elmtheory}

Extreme learning machines (ELMs)~\citep{huang2006extreme} are a type of single-hidden-layer feedforward neural network designed to address the slow training speed of traditional gradient-based methods. Conventional training involves iterative tuning of network weights to minimize a loss function, usually taken as the error between the network's predictions and the training data output. In contrast, ELMs initiate the hidden-layer parameters randomly, freeze them during training, and determine the output weights analytically using a least-squares approach, which has only a fraction of the computational cost of iterative methods. 

Initially, a dataset $\mathcal{D} = \{ \bm{X}, \bm{y} \}$ is assumed with $\bm{X} \in \mathbb{R}^{N_t \times N_d}$ and $\bm{y} \in \mathbb{R}^{N_t \times 1}$, with $N_d$ being the dimensions in each training input $\bm{x}_i$, for $i = 1, \dots, N_t$. The target $y_i$ is assumed to be one-dimensional, although extensions to multidimensional cases are straightforward. An ELM model calculates its outputs $o_i$ in a similar manner as a standard neural network, as

\begin{equation}
o_i = \bm{H}_i \bm{\beta},
\label{eq:elmoutput}
\end{equation}

\noindent where $\bm{H}_i \in \mathbb{R}^{1 \times N_n}$ is a matrix of $N_n$ hidden neural values calculated for the $i^{\mathrm{th}}$ input and $\bm{\beta}  \in \mathbb{R}^{N_n \times 1}$ is a weight matrix connecting the hidden nodes to the outputs. The hidden nodes $\bm{H}_i$ are obtained by

\begin{equation}
\bm{H}_i = f (\bm{x}_i \bm{W} + \bm{b}),
\label{eq:elmhidennodes}
\end{equation}

\noindent where $\bm{W} \in \mathbb{R}^{N_d \times N_n}$ is a weight matrix and $\bm{b} \in \mathbb{R}^{1 \times N_n}$ is a bias vector, typically consisting of random values sampled with $\bm{W},\bm{b} \sim \mathcal{N}(\bm{0},\bm{1})$, and $f$ is an activation function, responsible for accounting for non-linearities within the data. In the remainder of this work, the activation function adopted is the sigmoid, obtained as

\begin{equation}
f (x) = \frac{1}{1 + \mathrm{e}^{-x}}.
\label{eq:sigmoid}
\end{equation}

Training the ELM model consists of selecting optimal values in $\bm{\beta}$ that minimize the error between the outputs $\bm{o}$ and the dataset targets $\bm{y}$, such that 

\begin{equation}
\min_{\bm{\beta}} \sum_i^{N_t} \left\lVert \bm{H}_i \bm{\beta} - \bm{y}_i \right\rVert.
\label{eq:elmerror}
\end{equation}

\noindent Evaluating all inputs $\bm{X}$ yields the full hidden neural values matrix $\bm{H}$, which can be decomposed using singular value decomposition as

\begin{equation}
\bm{H} = \bm{U} \bm{\Sigma} \bm{V}^T,
\label{eq:elmsvd}
\end{equation}

\noindent allowing for the least-squares fit of the weight matrices via the generalized inverse operation

\begin{equation}
\bm{\hat{\beta}} = \bm{V} \bm{\Sigma}^{-1} \bm{U}^T \bm{y}.
\label{eq:elmbeta}
\end{equation}

\noindent With the optimal weight values, predictions are computed by calculating the hidden neural values $\bm{H}_{\star}$ with new inputs $\bm{X}_{\star}$, and further computing $\bm{o}_{\star} = \bm{H}_{\star} \bm{\hat{\beta}}$.

The least-squares method for evaluating $\bm{\hat{\beta}}$ (see Eq.~(\ref{eq:elmbeta})) can lead to overfitting in standard ELM models, especially when the number of neurons $N_n$ is large. To address this, regularization techniques, commonly used in standard neural networks, modify the loss function by adding terms that penalize large weight magnitudes. In regularized ELMs, this can be achieved through a Tikhonov regression approach~\citep{deng2009regularized,martinez2011regularized,zhang2020r} by including a penalty term $\lambda$, such that each $i^{\mathrm{th}}$ term in the inverse of the diagonal matrix $\bm{\Sigma}^{-1}$ is obtained as

\begin{equation}
\bm{\Sigma}_{i,i}^{-1} = \frac{\sigma_i}{\sigma_i^2 + \lambda^2}.
\label{eq:elmregul}
\end{equation}

\noindent Optimal values for the parameter $\lambda$ are generally obtained using cross-validation~\citep{browne2000cross,stone1978cross} or L-curves~\citep{hansen1992analysis}. It is noted that setting $\lambda = 0$ returns the model to the ordinary least squares of Eq.~(\ref{eq:elmbeta}). The algorithm for training and predicting with ELM models is shown in Alg~\ref{alg:elm}.

\begin{algorithm}[!htb]
    \caption{Extreme learning machines}
    \small
    \begin{algorithmic}[0]
        \State{\textbf{Inputs}}
        \State{\hspace{0.5cm}Training data $\mathcal{D} = \{ \bm{X}, \bm{y} \}$, number of neurons $N_n$}, regularization parameter $\lambda$
        \State{\textbf{Training}}
        \State{\hspace{0.5cm}Sample weights and biases from $\bm{W},\bm{b} \sim \mathcal{N}(\bm{0},\bm{1})$}
        \State{\hspace{0.5cm}Calculate the hidden layer output matrix $\bm{H} = f (\bm{X} \bm{W} + \bm{b})$}
        \State{\hspace{0.5cm}Compute weight matrix $\bm{\hat{\beta}}$}
        \State{\textbf{Prediction}}
        \State{\hspace{0.5cm}With test inputs $\bm{X}_{\star}$}
        \State{\hspace{0.5cm}Compute new weight matrix $\bm{H}_{\star} = f (\bm{X}_{\star} \bm{W} + \bm{b})$}
        \State{\hspace{0.5cm}Predict new outputs using $\bm{o}_{\star} = \bm{H}_{\star}\bm{\hat{\beta}}$}
    \end{algorithmic}
    \label{alg:elm}
\end{algorithm}

\subsection{ELM application to self-excited forces in bridge aerodynamics} \label{sec:elmapplication}

\begin{figure}[!htb]
\centering
% PLEASE DO NOT SCALE THE FIGURES!
\includegraphics[]{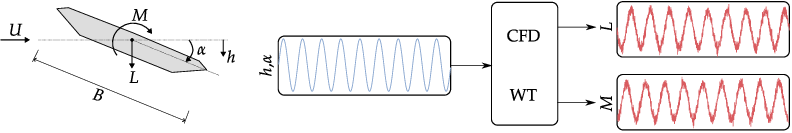}
\caption{Left: 2D coordinate system for the degrees of freedom ($h$ and $\alpha$) and the corresponding aerodynamic loads ($L$ and $M$), for a cross-section of width $B$ subjected to a flow field with mean wind speed $U$. Right: the general framework to obtain aerodynamic derivatives from aerodynamic analyses, where a harmonic motion is imposed to the system, and the loads are tracked either in computational fluid dynamics (CFD) or wind tunnel (WT) tests.}\label{fig:schematic}
\end{figure}

Self-excited forces in bridge decks arise from the interaction between the structural movement and the flow field. Assuming a 2D cross-section with width $B$ subjected to a flow field with mean wind speed $U$, any displacement in the vertical $h$ direction or rotations $\alpha$ around the shear centre generates aerodynamic lift $L$ and moment $M$ forces. A schematic of this consideration is shown in Fig.~\ref{fig:schematic} (left).

Early research on self-excited forces began with~\cite{theodorsen1979general}, who provided an analytical solution for self-excited loads ($L$ and $M$) on an idealized flat plate undergoing sinusoidal heaving and/or pitching in steady, incompressible flow. The plate was assumed to have zero thickness and a finite width. Improvements on the flat plate model were achieved by~\cite{scanlan1971airfoil}, who proposed a linear formulation for the aeroelastic given by

\begin{align}
L &= \frac{1}{2} \rho U^2 B \left( K H_1^* \frac{\Dot{h}}{U} + K H_2^* \frac{B \Dot{\alpha}}{U} + K^2 H_3^* \alpha + K^2 H_4^* \frac{h}{B}\right), \label{eq:scanlanL}\\
M &= \frac{1}{2} \rho U^2 B^2 \left( K A_1^* \frac{\Dot{h}}{U} + K A_2^* \frac{B \Dot{\alpha}}{U} + K^2 A_3^* \alpha + K^2 A_4^* \frac{h}{B}\right), \label{eq:scanlanM}
\end{align}

\noindent where $\rho$ is the air density and $K = B \omega / U$ is a reduced circular frequency of motion given an oscillation frequency $\omega = 2 \pi f$. The dimensionless aerodynamic derivatives $H_i^*$ and $A_i^*$ are cross-sectional properties given as functions of the reduced frequency $K$, and can be typically obtained for individual harmonic frequencies through CFD analysis or wind tunnel tests using scaled section models (see Figure~\ref{fig:schematic}, right). It is noted that extended formulations of the aerodynamic forces, including along-wind motion and drag effects, are available in the literature. While full ELM models incorporating these additional loads and degrees of freedom are feasible, they are not considered in this work for the sake of simplicity. Modern semi-analytical models for aerodynamic loads rely on various simplifications of the actual physical phenomena, as extensively reviewed by~\cite{kavrakov2018synergistic}. From a categorical perspective, these models ultimately confirm that CFD remains the most accurate approach for modelling real flow behaviour, second only to wind tunnel experiments~\citep{kavrakov2019categorical}. The approach proposed in this work aims to enhance the accuracy of semi-analytical methods by training the extreme learning machine (ELM) models directly on CFD data, with straightforward extensibility to wind tunnel data. Unlike semi-analytical methods based on simplifications of the underlying phenomena, this approach allows for a direct transfer function from the input motion to the complex data generated by CFD or wind tunnel experiments, better representing the complex aerodynamic behaviour, while retaining computational efficiency. 

\begin{figure}[!htb]
\centering
% PLEASE DO NOT SCALE THE FIGURES!
\includegraphics[]{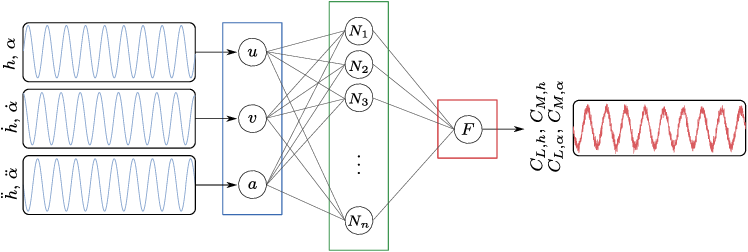}
\caption{Framework of the ELM models proposed in this work. The input layer (blue box) accepts the motion ($h$ or $\alpha$) and its time derivatives, represented by $u$, $v$ and $a$. The hidden layer (green box) contains a user-defined number of neurons $N_n$, used to calculate the output force $F$ (red box). In this approach, four ELM models are required to cover all combinations of input motion types and aerodynamic forces ($C_{i,j}$ for $i \in \{ L, M \}$ and $j \in \{ h, \alpha \}$).\label{ELM_Model_Schematic}}
\end{figure}

The ELM models adopted herein follow the structure illustrated in Figure~\ref{ELM_Model_Schematic}, where the inputs are either the vertical or rotational displacements and its time derivatives (represented in Figure~\ref{ELM_Model_Schematic} by $u$, $v$ and $a$), and the output ($F$ in Figure~\ref{ELM_Model_Schematic}) is the corresponding non-dimensional force value $C_{i,j}$ for $i \in \{ L, M \}$ and $j \in \{ h, \alpha \}$, calculated as
\begin{align}
C_{L,j} = \frac{L}{\frac{1}{2} \rho U^2 B}, \label{eq:CL}\\
C_{M,j} = \frac{M}{\frac{1}{2} \rho U^2 B^2}. \label{eq:CM}
\end{align}

\noindent In that manner, four different ELM models are required for each analysis case, covering all combinations of input motion types and aerodynamic forces.

\subsection{Design of experiments}

In the proposed framework, the training data consists of forced vibration datasets, where sinusoidal oscillations are applied to 2D section models, and the aerodynamic moment and lift forces are tracked over time. The oscillations have amplitudes of $\hat{h}$ for heave motion and $\hat{\alpha}$ for pitch, with the frequency $f$ governed by the non-dimensional reduced velocity

\begin{equation}
v_r = \frac{U}{Bf} = \frac{2 \pi}{K}.
\label{eq:vr}
\end{equation}

The selection of the number of oscillation cycles $N_c$ and the number of training points per cycle $N_{tp}$ depends on the origin and characteristics of the training data. For the analytical flat plate case in Sec.~\ref{sec:flatplate}, the forces are linearly related to the input motion and are noise-free, allowing for training with a single oscillation cycle. In contrast, the dataset for the Great Belt East case in Sec.~\ref{sec:bridgeaerodyn} is derived from CFD simulations, which are noisy and contain nonlinearities, especially at high angles of attack. Therefore, multiple oscillation cycles and more training points are necessary to capture the aerodynamic forcing behaviour more accurately. The selection of $N_{tp}$ is be guided by relating the sampling rate of the dataset to the frequency content of the nonlinear force components. Following the Nyquist theorem, the number of training points should be large enough to ensure that all relevant nonlinear components are adequately resolved in the training data. However, there is currently no best practice for the optimal selection of training data and model properties in bridge aerodynamics~\citep{abbas2020prediction}.

\begin{table}[!htb] 
\caption{Training properties and structural parameters used for the flat plate (FP, Sec.~\ref{sec:flatplate}) and the Great Belt East bridge (GB, Sec.~\ref{sec:bridgeaerodyn}) cases: Reynolds number $Re$, number of training oscillation cycles $N_c$, number of training points per cycle $N_{tp}$, minimum and maximum reduced velocities $v_{r,min}$ and $v_{r,max}$, maximum vertical and torsional amplitudes of motion $\hat{h}$ and $\hat{\alpha}$, structural vertical mass $m_h$ and torsional mass $m_{\alpha}$, bending frequency $f_h$ and torsional frequency $f_{\alpha}$ and damping ratio $\zeta$.\label{tab:trainingprops}}
\newcolumntype{C}{>{\centering\arraybackslash}X}
\begin{tabularx}{\textwidth}{CCCCCCCCCCCCC}
\toprule
\textbf{Sec} & $Re$ & $N_c$ & $N_{tp}$ & {$v_{r,\mathrm{min}}$} & {$v_{r,\mathrm{max}}$} & {$\hat{h}$} & {$\hat{\alpha}$} & $m_h$ & $m_{\alpha}$ & $f_h$ & $f_{\alpha}$ & $\zeta$\\
 & {\footnotesize [-]} & {\footnotesize [-]} & {\footnotesize [-]} & {\footnotesize [-]} & {\footnotesize [-]} & {\footnotesize [m]} & {\footnotesize [deg]} & {\footnotesize [t/m]} & {\footnotesize [tm$^2$/m]} & {\footnotesize [Hz]} & {\footnotesize [Hz]} & {\footnotesize [\%]} \\
\midrule
FP & - & 1 & 25 & 2 & 16 & 1.0 & 1.0 & 22.74 & 2470 & 0.100 & 0.278 & 1.0 \\
GB & $10^5$ & 10 & 250 & 2 & 16 & 5.0 & 12.0 & 22.74 & 2470 & 0.100 & 0.278 & 0.5 \\
\bottomrule
\end{tabularx}
\end{table}

Table~\ref{tab:trainingprops} presents the selected training parameters for each structural case studied in Sections~\ref{sec:flatplate}~and~\ref{sec:bridgeaerodyn}, along with the respective structural properties~\citep{larsen1993aerodynamic,kavrakov2018synergistic}: vertical mass $m_h$, torsional mass $m_{\alpha}$, vertical and torsional frequencies $f_h$ and $f_{\alpha}$, and damping ratio $\zeta$. For the Great Belt case, the Reynolds number is set to $Re = 10^5$ to reproduce realistic physical conditions. In contrast, the flat plate case does not directly involve a specific Reynolds number, as it is based on Theodorsen’s theory, which assumes potential flow. The ranges of reduced velocity, $v_{r,\mathrm{min}}$ to $v_{r,\mathrm{max}}$, as well as the maximum motion amplitudes $\hat{h}$ and $\hat{\alpha}$, are selected to encompass the expected frequencies and amplitudes of oscillation for the respective models. Although these parameters are not directly related to the ELM model itself, they are essential for ensuring a physically consistent representation of the structural model from which the training data are generated. As in any machine learning approach, the quality of the predictions is directly dependent on the quality of the training data.

The selection of the normalisation parameter $\lambda$ and the number of neurons $N_n$ depends primarily on the complexity of the dataset and the level of noise present in the data. To determine suitable values, a cross-validation procedure is employed. The dataset is partitioned into $k$ subsets, where $k-1$ subsets are used for training and the remaining subset is reserved for testing~\mbox{\citep{jung2018multiple,rodriguez2009sensitivity}}. This process is repeated until all subsets have served as the testing set. For each partition, candidate values of $\lambda$ and $N_n$ are drawn from a large predefined range of possible values, 

\begin{align*}
10^{-10} \leq \lambda \leq 10^0, \\
10 \leq N_n \leq 5000,
\end{align*}

and the mean squared error between model predictions and testing data is recorded. The optimal parameters are then identified as those minimising the prediction error, with preference given to smaller $N_n$ values in order to reduce model complexity. Although this selection strategy is generally computationally demanding, it is feasible in the present case due to the high efficiency of the ELM algorithm (see Sec.~{\ref{sec:comp_performance}}).

Several comparison metrics will be used to evaluate the accuracy of the ELM models compared to analytical or CFD results~\citep{kavrakov2020comparison}. They assess both global and local signal features, including root mean square difference $\mathcal{M}_{\mathrm{rms}}$, correlation $\mathcal{M}_{\mathrm{c}}$, coefficient of determination $\mathcal{M}_{R^2}$, wavelet coefficients $\mathcal{M}_{\mathrm{W}}$, peak amplitudes $\mathcal{M}_{\mathrm{peak}}$, phase angles $\mathcal{M}_{\phi}$, and warped magnitude $\mathcal{M}_{\mathrm{m}}$. These metrics offer a normalized resemblance assessment, where values close to one indicate similar signals, while values near zero indicate dissimilarity.

\section{Analytical flat plate benchmark case} \label{sec:flatplate}

\subsection{Model training}

\begin{figure}[!htb]
\centering
% PLEASE DO NOT SCALE THE FIGURES!
\includegraphics[]{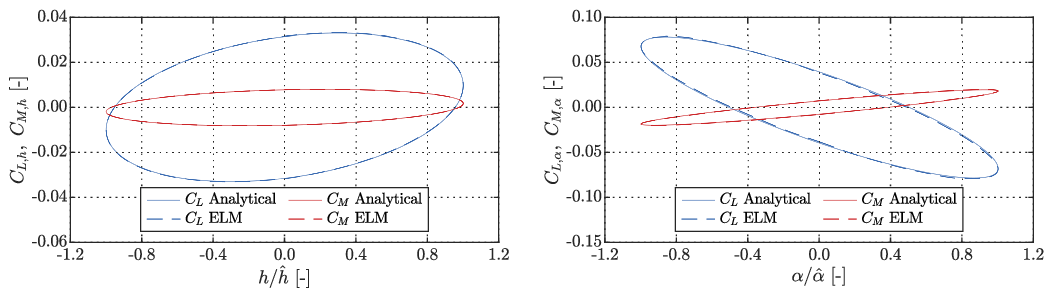}
\caption{Analytical flat plate case: prediction of normalized lift and moment forces in vertical (left, $C_{L,h}$ and $C_{M,h}$) and rotational (right, $C_{L,\alpha}$ and $C_{M,\alpha}$) directions due to a harmonic input motion, for a reduced velocity $v_r = 5$.\label{s001_FlatPlatePerformance}}
\end{figure}

To assess the ELM model's accuracy in predicting self-excited loads, it is first tested on the analytical case of a flat plate~\citep{theodorsen1979general}. In this scenario, aerodynamic lift and moment forces resulting from harmonic input motion are calculated analytically, assuming potential flow conditions. Since the model is linear, a single oscillation cycle is sufficient to represent aerodynamic behaviour. Each cycle is defined by a vertical amplitude of $\hat{h} = 1.0$~m and a rotational amplitude of $\hat{\alpha} = 1.0$~deg, with oscillation frequencies controlled by reduced velocities ranging from $v_r = 2$ to $v_r = 16$ at steps of $\Delta v_r = 2$ (see Eq.~(\ref{eq:vr}) and Table~\ref{tab:trainingprops}). Notably, for each oscillation cycle used in training, regardless of its specific frequency, the same number of training points $N_{tp}$ is used to ensure consistent data input across all oscillation cases. Since the training data is fully analytical and noise-free, the Tikhonov regularization parameter is set to $\lambda = 0$, and cross-validation determined the optimal number of neurons to be $N_n = 210$. Figure~\ref{s001_FlatPlatePerformance} shows a sample of the analytical results and ELM force predictions for the case of $v_r = 5$.

\subsection{Predictions of aerodynamic derivatives and random motion forces}

\begin{figure}[!htb]
\centering
% PLEASE DO NOT SCALE THE FIGURES!
\includegraphics[]{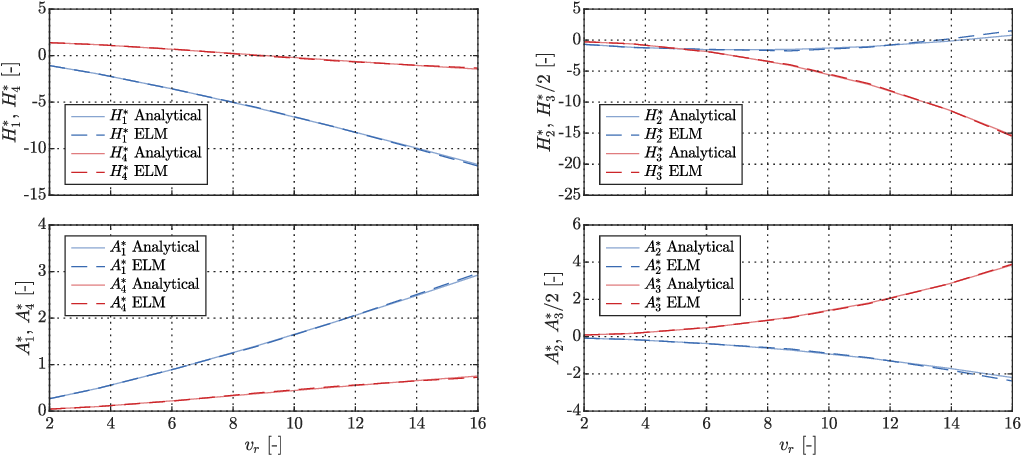}
\caption{Analytical flat plate case: comparison of aerodynamic derivatives obtained analytically and from the ELM model, estimated via least squares fit to Scanlan's formulation (see Eqs.~(\ref{eq:scanlanL},\ref{eq:scanlanM})).\label{s001_FlatPlateDerivatives}}
\end{figure}

Aerodynamic self-excited forces are typically represented by aerodynamic derivatives, which are understood as additional stiffness and damping terms resulting from fluid-structure interaction~\citep{scanlan1971airfoil}. For the flat plate case, the derivatives have an analytical, closed-form solution. Alternatively, they can be obtained using a least-squares fit, relating the input harmonic motion to the frequency-equivalent force content. Figure~\ref{s001_FlatPlateDerivatives} compares the aerodynamic derivatives obtained analytically with those derived from ELM-predicted forces. Overall, the ELM reaches a good agreement with the analytical counterpart model, with minor discrepancies appearing at high $v_r$ values, where the forces approach a quasi-steady state.

\begin{figure}[!htb]
\centering
% PLEASE DO NOT SCALE THE FIGURES!
\includegraphics[]{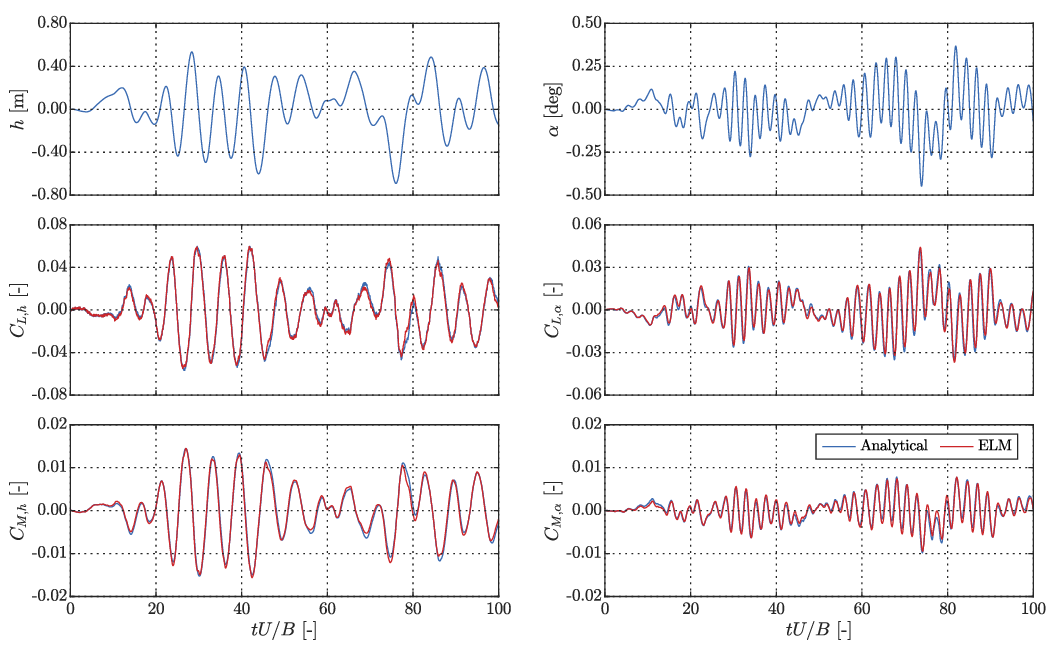}
\caption{Analytical flat plate case: a pseudo-random motion is applied to the structural model in both vertical (left) and rotational (right) degrees of freedom (top). The aerodynamic lift (centre) and moment (bottom) predicted by the trained ELM models are compared with the analytical solutions for both degrees of freedom.\label{s001_FlatPlateRandomMotion}}
\end{figure}

The true potential of the ELM model, trained using forces derived from harmonic motion input, lies in its ability to interpolate predictions to new frequency components not included in the training data. To demonstrate this capability, a pseudo-random input motion is generated for both vertical and rotational directions (see Figure~\ref{s001_FlatPlateRandomMotion}, top). The trained ELM models, developed using only harmonic motion data, are then used to predict the equivalent forcing signals. For the analytical flat plate case, these forces can be obtained by convolving the input motion with Wagner's indicial function~\citep{wagner1925}, which corresponds to the aerodynamic derivatives in the time domain. ELM predictions of the aerodynamic forces due to the pseudo-random input motion (Figure~\ref{s001_FlatPlateRandomMotion}, centre for lift and bottom for moment) are in good agreement with the analytical model, for cases associated with both vertical and rotational motions.

\begin{figure}[!htb]
\centering
% PLEASE DO NOT SCALE THE FIGURES!
\includegraphics[]{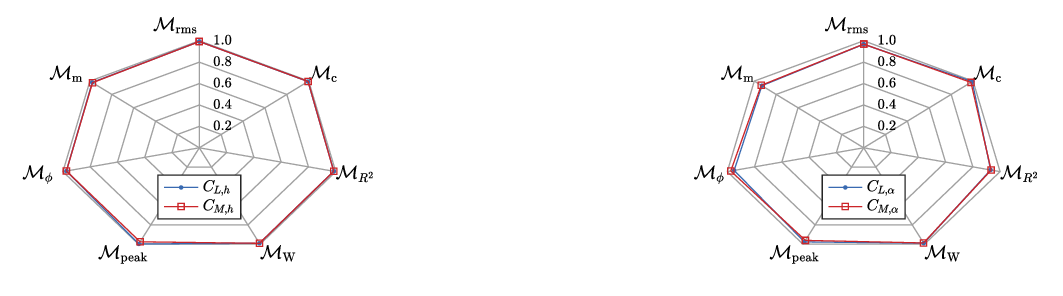}
\caption{Analytical flat plate case: comparison metrics related to root mean squared error $\mathcal{M}_{\mathrm{rms}}$, correlation $\mathcal{M}_{\mathrm{c}}$, coefficient of determination $\mathcal{M}_{R^2}$, wavelet $\mathcal{M}_{\mathrm{W}}$, peak amplitude $\mathcal{M}_{\mathrm{peak}}$, phase angle $\mathcal{M}_{\phi}$ and magnitude $\mathcal{M}_{\mathrm{m}}$. The results for both lift $C_L$ and moment $C_M$ forces are shown for the cases of vertical displacements (left) and rotations (right).\label{s001_FlatPlateComparisonMetrics}}
\end{figure}

To evaluate the ELM model predictions against the analytical solution, additional comparison metrics are employed (c.f.~\cite{kavrakov2020comparison}) and presented in Figure~\ref{s001_FlatPlateComparisonMetrics} (left, forces from vertical motion, and right, forces from rotational motion). Overall, the lift and moment forces show similar results. However, greater differences are observed when comparing forces from vertical and rotational motions, particularly in magnitude $\mathcal{M}_{\mathrm{m}}$, phase $\mathcal{M}_{\phi}$, and $\mathcal{M}_{R^2}$ metrics. The predictions for vertical motion are generally more accurate than those for rotational motion. These discrepancies arise from differences in signal norm, relative phase angles across frequency components, and overall prediction error relative to true forces. The evaluation of root mean squared error $\mathcal{M}_{\mathrm{rms}}$, correlation $\mathcal{M}_{\mathrm{c}}$, wavelet magnitude $\mathcal{M}_{\mathrm{W}}$, and peak amplitude $\mathcal{M}_{\mathrm{peak}}$ shows strong agreement with the analytically calculated forces. In summary, all metrics are close to unity, confirming the agreement observed qualitatively in Figure~\ref{s001_FlatPlateRandomMotion}.

\section{Application in bridge aerodynamics} \label{sec:bridgeaerodyn}

\subsection{Model training}

\begin{figure}[!htb]
\centering
% PLEASE DO NOT SCALE THE FIGURES!
\includegraphics[]{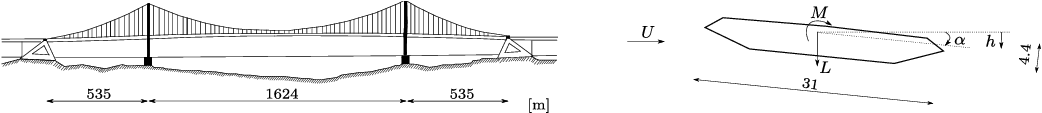}
\caption{Great Belt East Bridge schematic. Left: a side view of the bridge, showing the main and side spans. Right: the section dimensions, wind speed direction, coordinate system and aerodynamic forces.\label{GBSchematic}}
\end{figure}

\begin{figure}[!htb]
\centering
% PLEASE DO NOT SCALE THE FIGURES!
\includegraphics[]{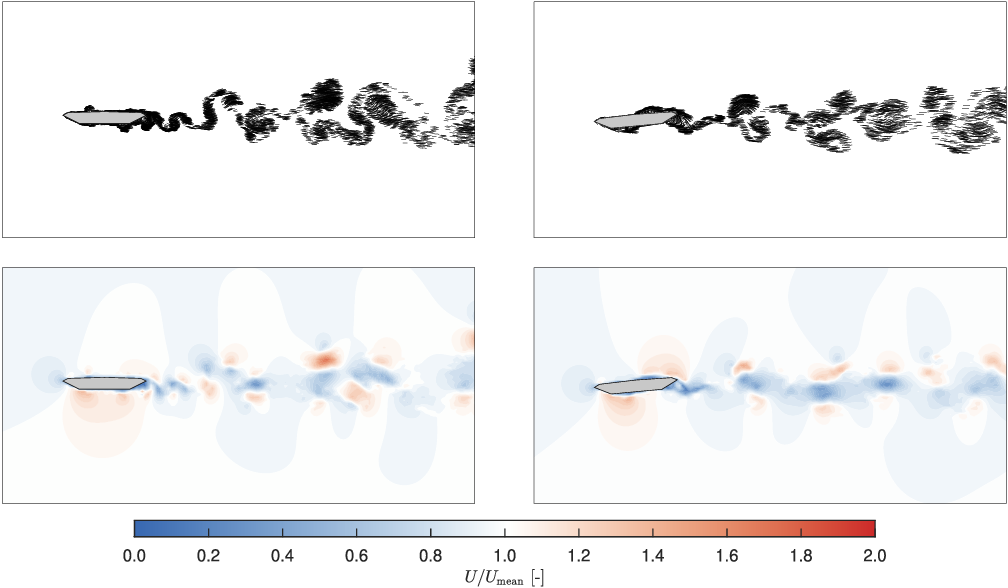}
\caption{Great Belt East Bridge: forced vibration simulations ($v_r = 2$ and $\hat{\alpha} = 5$~deg) from CFD analysis used as training data, in vertical (left) and rotational (right) directions. Vortex particle map (top) at $tU/B = 9.66$, and the respective normalised velocity fields (bottom). \label{VXFlowSim}}
\end{figure}

The ELM framework is applied to the Great Belt East Bridge model (see Figure~\ref{GBSchematic}). In this application, the focus is on a 2D model of the bridge deck cross-section. However, the general concept can be readily extended to full structural analysis by training separate ELM models for the relevant cross-sections with their respective datasets, and then performing the analysis e.g. in modal space. The aerodynamic forces in this application were calculated numerically using the Vortex Particle Method (VPM), with an in-house CFD solver thoroughly validated for different applications (see e.g.~\citep{morgenthal2002aerodynamic, KAVRAKOV2018825, TESFAYE2022103680, KAVRAKOV2019103971, chawdhury2018numerical, CHAWDHURY2021104391}). Snapshots of the particle map and the velocity fields for the force motion simulations in both vertical and rotational directions are shown in Figure~\ref{VXFlowSim}. The training data consists of forced harmonic oscillations from CFD simulations, with reduced velocities $v_r$ ranging from 2 to 16 and amplitudes of $\hat{\alpha} = { 1.0, 3.0, 5.0, 7.5, 10.0, 12.0}$~deg for the rotational case. In the vertical motion, the amplitude of vibration is calculated as

\begin{equation}
\hat{h} = \mathrm{atan}(\hat{\alpha}) \frac{U}{2 \pi f}.
\label{eq:heaveamp}
\end{equation}

\noindent Additional model properties are provided in Table~\ref{tab:trainingprops}. For each combination of reduced velocity and amplitude, 2500 training points were used ($N_c = 10$ cycles, each with $N_{tp}=250$ points). The increased $N_{tp}$ values aim at identifying non-linear components in the force signal, represented by higher harmonic components. Using cross-validation, the optimal number of neurons was determined as $N_n = 1890$ and the Tikhonov regularization parameter as $\lambda = 3 \times 10^{-7}$. 

\begin{figure}[!htb]
\centering
% PLEASE DO NOT SCALE THE FIGURES!
\includegraphics[]{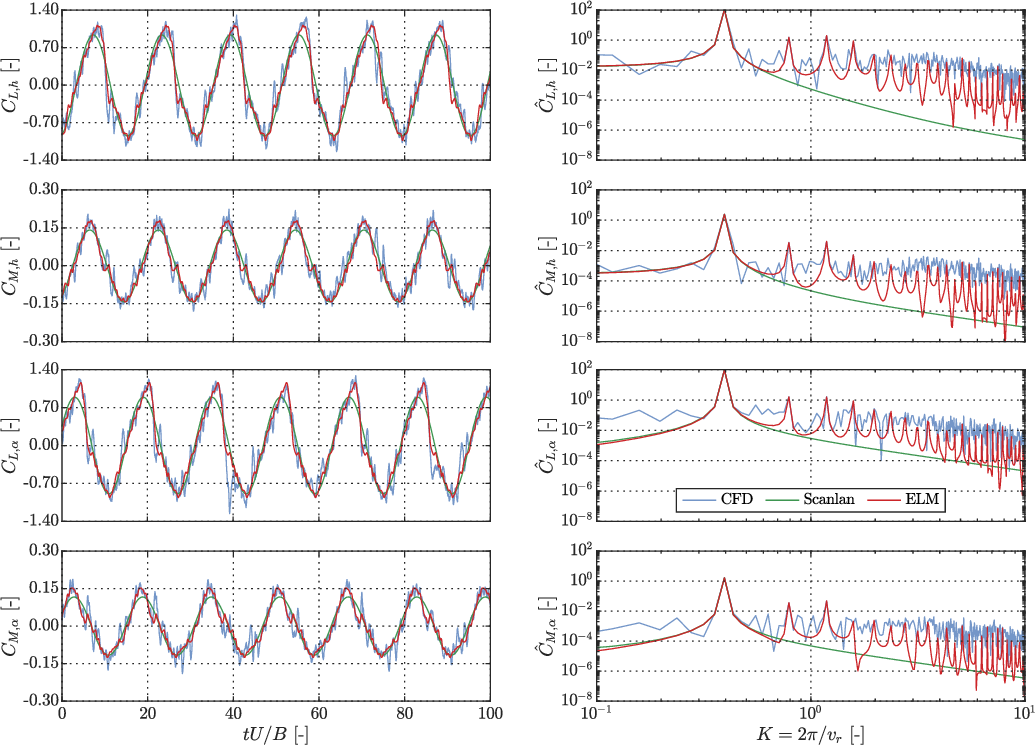}
\caption{Great Belt East Bridge: forced oscillation CFD simulations for an amplitude of $\hat{\alpha} = 10$~deg and a reduced velocity $v_r = 16$. Results for the normalized lift (top) and moment (centre-top) due to a vertical harmonic motion and lift (centre-bottom) and moment (bottom) due to a rotational harmonic motion are compared to the ELM predictions and the Scanlan linear fit. Comparisons are given in the time-domain normalized forces (left) and the corresponding power spectral density (right).\label{s002_GBPerformance}}
\end{figure}

Despite the streamlined geometry of the bridge deck, the CFD results for forced oscillation simulations of the Great Belt East Bridge model exhibit non-linear behaviour at high vibration amplitudes, as shown in Figure~\ref{s002_GBPerformance} for $\hat{\alpha} = 10$~deg and $v_r = 16$. After training, the ELM models predict forces induced by bridge motion in good agreement with the CFD results. Notably, the ELM accurately captures non-linear force components, represented by high-frequency peaks in the spectral amplitude at harmonics of the input motion frequency. This performance improves on the linear Scanlan model, also shown in Figure~\ref{s002_GBPerformance}, which captures only the primary frequency peak while neglecting non-linear contributions. 

\begin{figure}[!htb]
\centering
% PLEASE DO NOT SCALE THE FIGURES!
\includegraphics[]{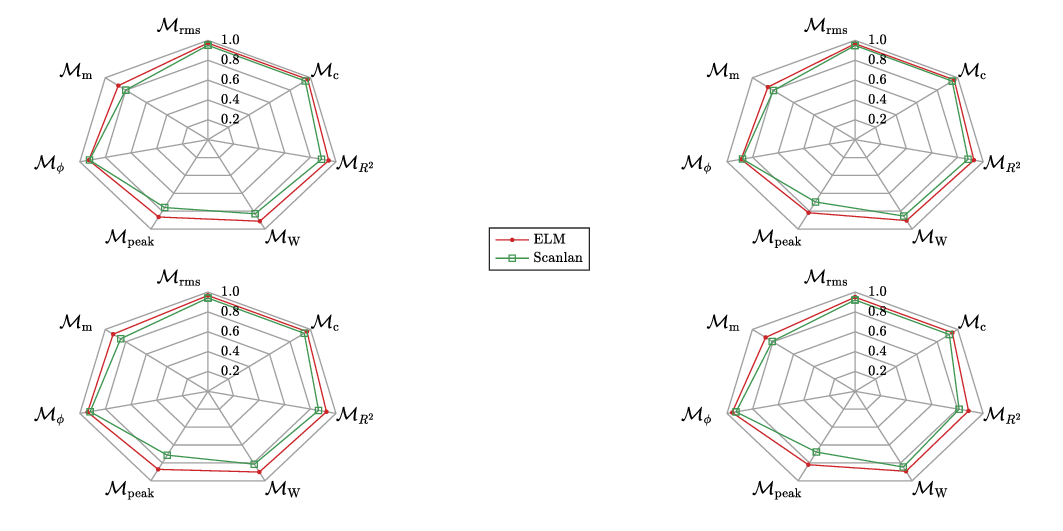}
\caption{Comparison metrics for the forced oscillation case of amplitude of $\hat{\alpha} = 10$~deg and a reduced velocity $v_r = 16$ (c.f. Figure~\ref{s002_GBPerformance}). The ELM model predictions are compared to Scanlan's linear fit, taking the CFD forces as a basis. Top: metrics for $C_{L,h}$ (left) and $C_{M,h}$ (right). Bottom: metrics for $C_{L,\alpha}$ (left) and $C_{M,\alpha}$ (right).\label{s002_GBForcedVibComparisonMetrics}}
\end{figure}

A detailed comparison of the ELM and Scanlan models for the forced oscillation results of Figure~\ref{s002_GBPerformance}, using CFD forces as a reference, is presented in Figure~\ref{s002_GBForcedVibComparisonMetrics}, with evaluation metrics provided for each force component. Due to the streamlined geometry of the deck section, the Scanlan model offers a reasonable approximation of the CFD forces, particularly for root mean square $\mathcal{M}_{\mathrm{rms}}$, correlation $\mathcal{M}_{\mathrm{c}}$, and phase angle $\mathcal{M}_{\phi}$, which are primarily influenced by the dominant frequency component. However, the ELM model shows significant improvements in magnitude $\mathcal{M}_{\mathrm{m}}$ and peak amplitude $\mathcal{M}_{\mathrm{peak}}$ metrics, also reflecting the influence of higher harmonics of the force magnitude. Additionally, the ELM outperforms the Scanlan model in the coefficient of determination $\mathcal{M}_{R^2}$, indicating a better statistical fit to the CFD data. The wavelet metric $\mathcal{M}_{\mathrm{w}}$ also favours the ELM, consistent with the improved frequency content representation observed in Figure~\ref{s002_GBPerformance}, where the ELM captures higher harmonic contributions.

\subsection{Prediction of aerodynamic derivatives and flutter instability}

\begin{figure}[!htb]
\centering
% PLEASE DO NOT SCALE THE FIGURES!
\includegraphics[]{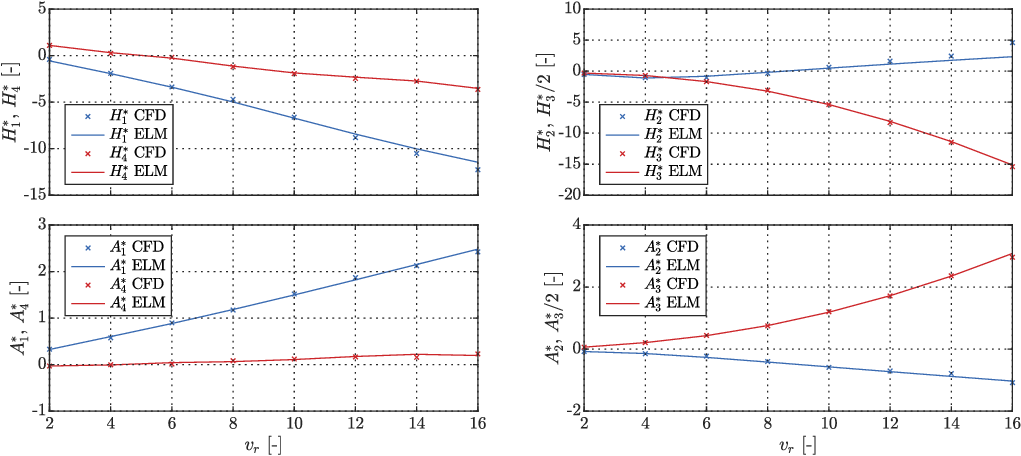}
\caption{Great Belt East Bridge: aerodynamic derivatives obtained from harmonic oscillation analysis computed both from CFD simulations and using the trained ELM models.\label{s002_GBDerivatives}}
\end{figure}

The trained ELM model is applied to estimate the aerodynamic derivatives of the Great Belt East Bridge by generating forced motion signals at various reduced velocities and fitting them to Scanlan's model using a least-squares approach. As shown in Figure~\ref{s002_GBDerivatives}, the ELM predictions align well with the CFD-derived results, also used as training data. The ELM model can reliably estimate derivatives for any $v_r$ value within the training range, eliminating the need for additional CFD simulations. 

\begin{figure}[!htb]
\centering
% PLEASE DO NOT SCALE THE FIGURES!
\includegraphics[]{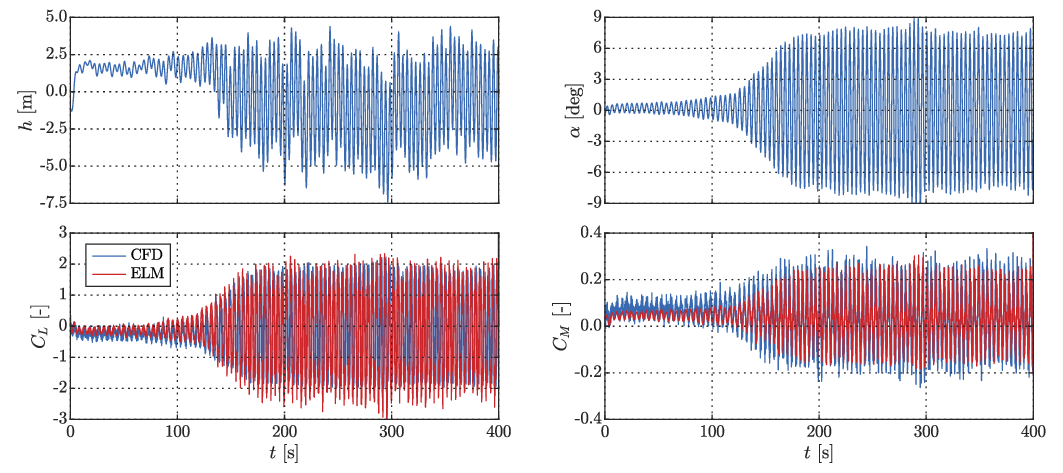}
\caption{Great Belt East Bridge: comparison of aerodynamic forces predicted by the ELM with the ones derived from a CFD analysis of flutter instability, showing limit cycle oscillation at $U=72.5$~m/s. Top: the vertical (left) and rotational (right) structural responses. Bottom: the lift (left) and moment (right) forces, linearly superimposed from each degree of freedom in the case of ELM predictions.\label{s002_GB_FlutterCFD_TimeDomain}}
\end{figure}

\begin{figure}[!htb]
\centering
% PLEASE DO NOT SCALE THE FIGURES!
\includegraphics[]{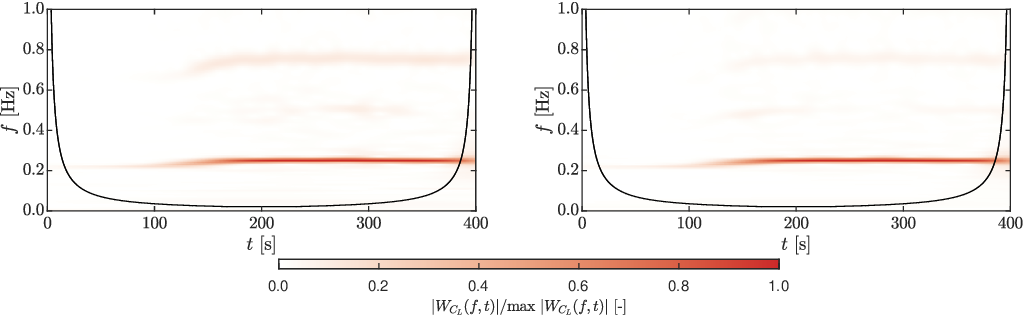}
\caption{Great Belt East Bridge: normalised Morlet wavelet analysis (with a central frequency $f_c = 4$~Hz) of the lift forces from a flutter simulation (see Figure~\ref{s002_GB_FlutterCFD_TimeDomain}), computed from CFD (left) and predicted by the ELM model (right).\label{s002_GB_CFD_Wavelet}}
\end{figure}

To assess the ELM model's performance under high-amplitude motion, a flutter analysis was conducted for the Great Belt East Bridge using CFD at a wind speed of $U = 72.5$~m/s. At this speed, the system exhibits divergent oscillation behaviour at the flutter onset, leading to a limit cycle oscillation with amplitudes of approximately $\hat{h} = \pm 3.75$~m and $\hat{\alpha} = \pm 6$~deg, as shown in Figure~\ref{s002_GB_FlutterCFD_TimeDomain} (top). These oscillation signals were input into the trained ELM model, and the predicted lift and moment forces were linearly superimposed and compared to the CFD-derived forces (Figure~\ref{s002_GB_FlutterCFD_TimeDomain}, bottom). The ELM model demonstrated accurate predictions of the high-amplitude self-excited forces, with slight overestimation of lift amplitudes and underestimation of moment loads. These discrepancies likely arise from both ELM prediction errors and the limitations of applying the linear superposition principle at high motion amplitudes. Importantly, the ELM model also accurately captures the frequency content of the coupled flutter behaviour, which lies between the structural natural frequencies (see Table~\ref{tab:trainingprops}), as shown in the wavelet analysis of the lift forces in Figure~\ref{s002_GB_CFD_Wavelet}. 

\begin{figure}[!htb]
\centering
% PLEASE DO NOT SCALE THE FIGURES!
\includegraphics[]{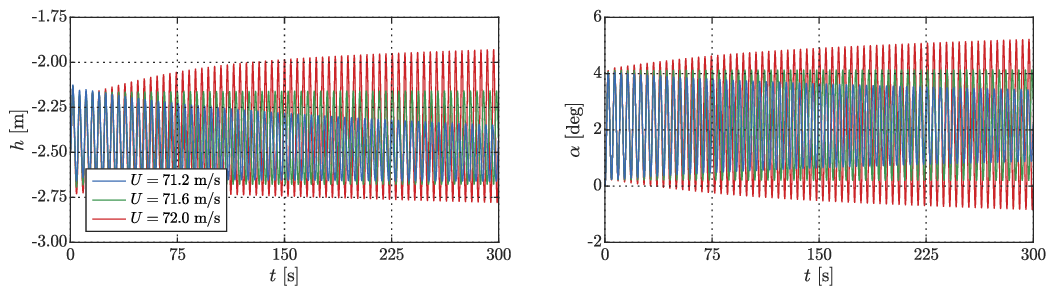}
\caption{Great Belt East Bridge: flutter analysis in time domain performed using a coupled ELM-structural model. Results are shown for the vertical displacements (left) and rotations (right), and indicate the onset of flutter at $U=71.6$~m/s, with higher wind speeds showing divergent response behaviour.\label{s002_GBFlutter}}
\end{figure}

\begin{figure}[!htb]
\centering
% PLEASE DO NOT SCALE THE FIGURES!
\includegraphics[]{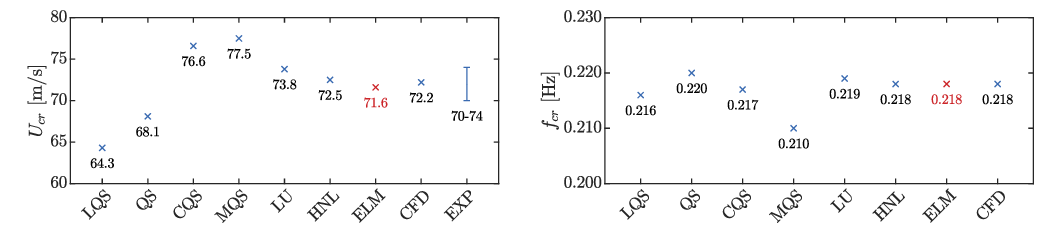}
\caption{Great Belt East Bridge: ELM flutter limit wind speeds (left) and the corresponding oscillation frequency of the coupled motion (right) in comparison with the linear quasi-steady (LQS), quasi-steady (QS), corrected quasi-steady (CQS), modified quasi-steady (MQS), linear unsteady (LU) and hybrid nonlinear (HNL) semi-analytical models, as well as CFD and wind tunnel results~\citep{kavrakov2018synergistic,larsen1993aerodynamic}.\label{s002_GBFlutterProps}}
\end{figure}

The trained ELM models are integrated with a structural model to estimate the flutter limit speed through time-domain analysis. Results in Figure~\ref{s002_GBFlutter} indicate the onset of flutter at a wind speed of $U = 71.6$~m/s. Simulations at lower wind speeds show convergent behaviour with decaying amplitudes, while higher wind speeds exhibit divergent oscillations. The limit cycle oscillation observed in CFD simulations (Figure~\ref{s002_GB_FlutterCFD_TimeDomain}) could not be replicated by the ELM models, likely due to the model structure and characteristics of the training data. Nevertheless, the ELM framework offers significant improvements over several semi-analytical methods often used for their simplicity and computational efficiency, as shown in Figure~\ref{s002_GBFlutterProps}. Compared to CFD results, the ELM models achieve a flutter limit wind speed with only a 0.8\% error and match the oscillation frequency almost exactly. In contrast, semi-analytical methods such as the linear quasi-steady (LQS), quasi-steady (QS), corrected quasi-steady (CQS), and modified quasi-steady (MQS) models exhibit errors exceeding 5\%~\citep{kavrakov2018synergistic}. The linear unsteady (LU) model, based on aerodynamic derivatives, has a 2.2\% error, underscoring the importance of non-linear force components in flutter simulations. Among the methods evaluated, only the hybrid nonlinear (HNL) model outperforms the ELM framework. The approach of the HNL model combines the nonlinear QS model for low reduced frequencies, where quasi-steady assumptions are valid, with the LU formulation for high reduced frequencies, where fluid memory effects dominate. Compared to wind tunnel results~\citep{larsen1993aerodynamic}, however, the LU, HNL, CFD and ELM models are in the range considered critical for the onset of flutter.

\subsection{Computational performance} \label{sec:comp_performance}

The ELM models were implemented in Python using standard open-source numerical packages. All simulations were executed on a desktop workstation equipped with an Intel~Core~i7~processor (3.0~GHz, 8~cores) and 8~GB of RAM under Windows 10. No graphics processing units (GPUs) or specialised hardware were used. Training times quoted in this paper include both the random initialisation of hidden weights and the singular value decomposition used to compute the output weights. The ELM model implementation is available in \url{https://github.com/gledsonrt/SelfExcitedELMs}.

The computational efficiency of ELMs is evident when compared to similar machine learning models. The ELM model completed training in 7.6 seconds in the Great Belt East Bridge example. In contrast, using the same number of neurons $N_n$, a single-layer feedforward neural network with a comparable RMS error target required 665.2 seconds, as it relies on gradient descent for training. Gaussian process regression took over $10^4$ seconds to train due to its non-parametric nature, as training time depends on the dataset size. The remarkable efficiency of ELMs enables their training to be performed within reasonable times on standard computers, eliminating the need for specialized machine learning hardware and associated costs.

The data used to train ELM models for aeroelastic forces is obtained from forced vibration simulations, which can be performed quickly and efficiently using CFD or wind tunnel tests. Once trained, ELM models can simulate a wide range of aeroelastic scenarios, particularly when coupled with structural models (see e.g. Figures~\ref{s002_GB_FlutterCFD_TimeDomain} and~\ref{s002_GBFlutter}). In time-domain flutter simulations for the Great Belt East Bridge, ELM predictions required an average computation time of just 6.2 seconds, a total of $0.0081$\% of the $7.7 \times 10^4$ seconds needed for CFD simulations on the same hardware. This enables rapid and reliable exploration of various structural analysis scenarios, making ELMs particularly valuable during the early stages of bridge design, or for applications in structural optimisation.

\section{Conclusion} \label{sec:conclusion}

This paper presented a methodology for modelling aerodynamic self-excited forces and instabilities using a data-driven approach based on extreme learning machines (ELMs), a class of single-layer feedforward neural networks that utilize matrix pseudo-inversion techniques to significantly accelerate the training process.

The proposed framework leverages forced oscillation datasets for training, enabling the systematic selection of training data from CFD simulations or wind tunnel experiments. This approach is particularly valuable as it offers guidance on selecting suitable datasets for machine learning models in bridge aerodynamics. Moreover, such tests are easily reproducible in early stages of design, as they do not require extensive wind tunnel campaigns or complex CFD simulations, while still capturing nonlinearities in aerodynamic forces arising from fluid-structure interactions.

Extreme learning machines were chosen for their simplicity and efficiency. They achieve an accuracy comparable to standard neural networks of similar size and properties but require a fraction of the training time. This facilitates the selection of optimal model parameters, obtained in this work through K-fold cross-validation. Once trained, ELM models predict aerodynamic loads in a fraction of the time needed for CFD simulations or wind tunnel experiments, enabling fast and accurate aerodynamic analysis of structures.

The performance of the ELM models was evaluated against several semi-analytical models from the literature. The results showed improved accuracy, particularly in capturing key aerodynamic behaviours and nonlinearities that are often overlooked in standard semi-analytical approaches. In comparison with CFD or wind tunnel tests, the ELM models provided reliable predictions of aerodynamic loads and their influence on flutter instability, accurately identifying both the onset wind speed and the coupled frequency content. This demonstrates that complex simulation results, typically requiring costly wind tunnel campaigns or computationally intensive CFD simulations, can instead be obtained from simple early-stage forced vibration tests coupled with the ELM model.

In summary, ELM models offer a fast and accurate alternative to CFD and wind tunnel simulations for modelling aerodynamic self-excited forces. They can be seamlessly integrated into the initial analysis and design processes for long-span bridges, providing a practical and efficient tool for aerodynamic evaluation. Future research may extend the ELM framework to 3D structures, where different cross-sections can be combined for more comprehensive load modelling. This may also include incorporating along-wind motions and drag effects into the ELM force formulations, as well as addressing other types of aerodynamic loads, such as buffeting, where gusty turbulent wind conditions dominate the response.

\bibliographystyle{arxiv3} 
\bibliography{references.bib}

\end{document}